\documentclass[compsoc, conference, a4paper, 10pt, times]{IEEEtran}

\usepackage{cite}
\usepackage{amsmath,amssymb,amsfonts}
\usepackage{algorithmic}
\usepackage{graphicx}
\usepackage{textcomp}
\usepackage{xcolor}
\usepackage{tabularx}
\usepackage{booktabs}
\usepackage{cleveref}
\usepackage{csquotes}

\usepackage{tikz}
\usepackage{subcaption}
\usepackage{url}
\usepackage{array}
\usepackage{xspace}

\newcommand{\hLine}[3]{\begin{tikzpicture}
		[overlay,color=gray,line width=2pt]
		\draw (#1,#2) -- (#3,#2);
	\end{tikzpicture}
}

\begin{document}

\title{Case Study: Estimating the Differential Identifiability Risk of an SOEP Data Set}
\title{An applied Perspective: Estimating the Differential Identifiability Risk\\ of an Exemplary SOEP Data Set\vspace{-1em}}

\author{
\IEEEauthorblockN{1\textsuperscript{st} Allmann, Jonas}
\IEEEauthorblockA{\textit{Technische Universit\"at Berlin} \\
j.allmann@campus.tu-berlin.de}
\and
\IEEEauthorblockN{2\textsuperscript{nd} Nu{\~n}ez von Voigt, Saskia}
\IEEEauthorblockA{\textit{Technische Universit\"at Berlin} \\
saskia.nunezvonvoigt@tu-berlin.de}
\and
\IEEEauthorblockN{3\textsuperscript{rd} Tschorsch, Florian}
\IEEEauthorblockA{\textit{Technische Universit\"at Dresden} \\
florian.tschorsch@tu-dresden.de}
}

\maketitle

\begin{abstract}
  Using real-world study data usually requires contractual agreements where research results may only be published in anonymized form.
  Requiring formal privacy guarantees, such as differential privacy,
  could be helpful for data-driven projects to comply with data protection.
  However, deploying differential privacy in consumer
  use cases raises the need to explain its underlying mechanisms and the resulting
  privacy guarantees.
  In this paper, we thoroughly review and extend an existing privacy metric.
We show how to compute this risk metric efficiently for a set of basic statistical queries.
  Our empirical analysis based on an extensive, real-world scientific
  data set
expands the knowledge on how to compute
  risks under realistic conditions, while presenting more challenges than
  solutions.
\end{abstract}

\section{Introduction}
As part of the German Socio-Economic Panel Study~(SOEP),
tens of thousands of individuals in thousands of households have been and are
surveyed each year.
The SOEP is the largest and longest-running multidisciplinary long-term study in Germany.
The study is conducted annually since 1986 by the German Institute for Economic Research
\textit{DIW}~\cite{goebel_german_2019,berlin_diw_nodate}.
Even under the assumption that the data is handled properly by every single
research team and the raw data set is never leaked outside authorized circles,
aggregate statistics computed from it are regularly released as part of
SOEP-based research publications.
However, even precise aggregate statistics can pose a privacy risk~\cite{dwork_differential_2006}.

Differential privacy~\cite{dwork_differential_2006} is a prominent privacy framework to
technically and verifiably protect the data from re-identification attacks
in published statistics.
The main concern with differential privacy is the trade-off between
data utility and privacy. It is generally easier to achieve
strong privacy guarantees (by choosing a small~$\varepsilon$) in combination
with high data utility when a data set is large, making the SOEP
an interesting candidate for a meaningful implementation of
differential privacy.

Even though the optimization of the privacy-utility trade-off is still a major
challenge, we focus on
how to effectively
communicate this notion of privacy protection to participants.
There remain risks to privacy even after its implementation, 
and the guarantees it provides are not trivial to understand, 
even for trained professionals~\cite{nanayakkara_visualizing_2022}.
Differential privacy relies on statistical and probabilistic 
mechanisms and its precise guarantees depend on configurable 
parameters, as well as on the data being processed and the type
of analysis being performed.

While the overall interpretation of the privacy guarantee provided by
differential privacy is intuitive, the risk implications 
of specific values of $\varepsilon$ are more challenging to understand. 
One approach to making the privacy parameter of differential privacy more
comprehensible is to translate $\varepsilon$ into a corresponding privacy risk,
expressed as a risk of individual identifiability~\cite{lee_how_2011,mehner_towards_2021}.
This privacy risk metric depends not only on
$\varepsilon$, but also on the data that is processed and the statistical query~\cite{lee_how_2011}.
However, it is not clear how such risk metrics
perform under real-world conditions, which is a crucial step towards making risk
communication formats viable for practical applications.

The major goal of this paper is to look at the concept of the privacy risk metric from an applied perspective.
We investigate the interaction of the risk metric
of Lee and Clifton~\cite{lee_how_2011} with an SOEP data set.
Our results show that precise risk calculations depend heavily on the actual data involved,
on extreme values and, not least, on the number of individuals included in a
database and in specific query results. To extend the current knowledge of how
practical risk metrics for differential privacy are under real world
conditions, it seems necessary to intensify the investigation of these aspects.

The paper's contribution and structure can be summarized as follows:
In \Cref{sec:background}, we provide
essential knowledge about
differential privacy and its core concepts, and present
an existing risk metric published by Lee and Clifton.
In \Cref{sec:theoreticalAnalysis}, we extend the risk metric to a wider range of query types and
show how to calculate the risk for an SOEP data set.
We present the results of our empirical evaluation in
\Cref{sec:evaluation}.
We summarize the challenges that remain by calculating the differential privacy risk for a real world data set in
\Cref{sec:discussion}.
Lastly, we review related work in \Cref{sec:related_work} and conclude our paper
with a brief outlook in \Cref{sec:conclusion}.

\section{Differential Privacy and Risk Model}
\label{sec:background}
Our research is based on the following context:
We assume that descriptive statistics of SOEP
variables are to be published as part of a data-driven
research project.
The sensitive nature of the SOEP data 
calls for guarantees like the ones differential privacy provides.
Using the Laplace mechanism, controlled noise is added to the query results
(cf. \Cref{sect:laplace}). The associated privacy risk
is published alongside the results, as we assume a scenario where
the data has already been collected anyway.
A challenge in this context is calculating this associated privacy risk.
To this end, we describe in \Cref{sect:risk_metrics}
risk metrics relying on an approach of Lee and Clifton~\cite{lee_how_2011}.

\subsection{Differential Privacy}
\label{sect:differential_privacy}

Differential privacy is supposed to protect individuals by making
any two databases differing in one individual statistically indistinguishable.
For potential data subjects, e.g. in the SOEP data set, that means that adding their data to the database, should not affect what can or cannot be learned
about them as an individual~\cite{dwork_differential_2006}.
A differentially private mechanism $M$ provides differential privacy for its
query result $q$, for all neighboring database instances $X$ and $X'$ that differ in one single element:
\begin{equation}
	\label{formula:differential_privacy}
	P[ M(X) = q ] \leq e^\varepsilon \cdot P[ M(X') = q ].
\end{equation}
$P$ is the probability operator and $\varepsilon$ is the
so-called privacy parameter. It balances privacy and utility in any
differential privacy implementation.

The parameter $\varepsilon$ captures the allowed distance between the
probabilistic output distributions of the mechanism $M$ when
applied to databases $X$ and $X'$. If $\varepsilon$ is small, the
tolerable difference in probability is also small, and
the privacy level is high.
A larger allowed distance between the output distributions,
makes the impact on query results of a single individual more
clearly discernible. This diminishes the level of privacy protection. At the
same time, utility of the output is increased, as less noise needs to be
added to the query result.

\subsubsection{Laplace Mechanism}
\label{sect:laplace}
Several mechanisms exist to achieve differential privacy in practice.
In the most common case, differences in raw query outputs are masked by a
calibrated amount of random noise, added before a query result is released.
For example, we add an amount of noise to the average hours spent of paid work
querying the SOEP data set.
The most commonly used source of noise is the Laplace distribution. It is
calibrated to the privacy parameter $\varepsilon$ and the global
sensitivity $\Delta f()$ of a query function $f$.

\subsubsection{Global Sensitivity}
The global sensitivity $\Delta f$ of a query function $f$ over a data universe
$U$ is defined as the largest possible difference in function value between two
neighboring inputs taken from $U$:
\begin{equation}
	\Delta f(U) = \max_{X, X' \subset U, h(X,X') = 1} (|f(X) - f(X')|).
	\label{formula:global_sensitivity}
\end{equation}

The global sensitivity can usually be
computed if the data universe is known, sometimes even without that condition.
A simple \textit{count} query, for example, always has a global
sensitivity of one, since adding or removing a single entry from a database can
only increase or decrease any \textit{count} by one.
For many other queries, global sensitivity under a known universe of values can be derived
analytically and it can be much larger than one for
other queries, especially in the presence of outliers.

\subsubsection{Local Sensitivity}
A major disadvantage of global
sensitivity is, depending on the data universe and the specific query
function, it often leads to the addition of much more noise than is actually
necessary to make a specific query differentially private.
To partially circumvent the disadvantages of global sensitivity, local
sensitivity was introduced. The central idea of local sensitivity is to take
into account the actual function value of a given data set or a given query
result and measure the distance in function value to any possible neighbour
data set. $\Delta v$ will be used as notation for the local sensitivity of a
query function $f$ from here on:
\begin{equation}
	\Delta v{(U, X)} = \max_{X, X' \subset U, h(X,X') = 1} (|f(X) - f(X')|).
	\label{formula:local_sensitivity}
\end{equation}

Here, additionally to the data universe $U$, local sensitivity depends on a
given subset $X \subset U$. Sensitivity is measured against any possible
neighbouring set $X' \subset U$. This approach can lead to drastically reduced
amounts of noise needed, since function sensitivity is often much smaller
locally than the global worst-case sensitivity~\cite{nissim_smooth_2007}.

In the contruction of the risk metrics presented in the next section,
global sensitivity is assumed as the reference parameter to calibrate noise,
while local sensitivity represents the potential influence of a data subject
on query results in a specific situation.

\subsection{Risk Metrics}
\label{sect:risk_metrics}
This section will turn to the
question of calculating risks for a given differentially private system.
We will rely heavily on an
approach by Lee and Clifton~\cite{lee_how_2011}. 
The risk metric of Lee and Clifton calculates, for
the Laplacian noise mechanism, a suitable $\varepsilon$ to guarantee that the risk
of a successful membership inference attack remains under a given threshold.
From here on, this metric will be referred to as $\rho_{mw}$ for many worlds,
a name that will become clear later. It is calculated as follows:
\begin{equation}
	\rho_{mw} \leq \frac{1}{1 + (n - 1) e^{-\frac{\varepsilon \Delta v}{\Delta f}}}.
	\label{formula:full_risk_metric}
\end{equation}
Here, $n$ denotes the total number of subjects in the data set at hand, $\Delta
v$ denotes the local sensitivity of the query function and $\Delta f$ its
global sensitivity. Local sensitivity should always be smaller or equal to
global sensitivity, and therefore $0 < \frac{\Delta v}{\Delta f} \leq 1$ should
always hold.
The $\varepsilon$, as usual, is the chosen privacy parameter and
$\rho_{mw}$ is the risk of being identified as present or absent in the
database.

When $n$ is large, $\rho_{mw}$ can become much smaller than~$0.5$.
The attacker of Lee and Clifton does not try to decide between
presence or absence of one individual. Instead, the number of individuals in a
database is known, and the attacker tries to find out, which individuals from a known
universe of data subjects are included. Finding the `true' database becomes a lot harder than
just deciding between presence or absence of one individual, and the obtained
tracing risk can be dramatically smaller than $0.5$.
Very broadly, the risk measure $\rho_{mw}$ is small when the number of subjects
$n$ is large, because more `possible worlds' need to be assumed by an
attacker. It is small when $\varepsilon$ is small, because a smaller $\varepsilon$
demands larger amounts of noise. And finally it is small when $\Delta v$ is
much smaller than $\Delta f$, because smaller local changes in function value
are hidden by the larger amounts of noise demanded by large global sensitivity.

The parameters needed to calculate $\rho_{mw}$ are not necessarily available in
every given scenario. Especially local and global sensitivity require knowledge
about the data universe and the data set at hand, that might be missing at the point in time
when the risk is calculated. In the context of our scenario, it would be impossible
to communicate a risk based on this metric to participants before any data is collected.
Precise risk computation and communication would only become possible after the participants'
decision to share their data has been made. This dilemma is addressed by modified
versions of the above metric.

Mehner et al.~\cite{mehner_towards_2021}
propose a simplified risk metric based on the
metric presented above. Their worst-case risk metric requires fewer parameters
to be calculated, resulting in reduced precision, but giving more flexibility
in practice.
Specifically, this approach avoids the assumption of knowing the
data universe in advance and being able to calculate precise values for global
and local sensitivity.
Assuming instead the worst case ratio of these two sensitivities
$(\Delta v/\Delta f=1)$ and
number of possible worlds ($n=2$) gives the maximum global privacy risk, referred to here as $\rho_{gl}$:
\begin{align}
	\rho_{gl} = \frac{1}{1 + e^{-\varepsilon}}.
	\label{formula:global_privacy_risk}
\end{align}

In contrast to the simplified metric proposed by Mehner
et al.~\cite{mehner_towards_2021}, in this paper we suggest
a new metric that keeps the fraction between local and
global sensitivity as an important factor influencing the risk for a
given data set. For the size of the data set it
assumes the worst-case parameter $n=2$, resulting in:
\begin{align}
	\rho_{tw} \leq \frac{1}{1 + e^{-\frac{\varepsilon \Delta v}{\Delta f}}}.
	\label{formula:two_worlds_risk}
\end{align}
Our newly proposed metric will be called the two-worlds metric
from here on, as opposed to the many-worlds metric that was proposed by Lee
and Clifton.

\section{Procedure for Risk Estimation}
\label{sec:theoreticalAnalysis}
It becomes clear that publishing differentially private statistics
with precisely bounded risks not only requires an adequate choice of $\varepsilon$,
but also detailed knowledge of sensitivities. The difficult task of choosing
$\varepsilon$ is usually focused on when investigating the privacy-utility trade-off.
Instead, we turn our attention the question of how sensitivities influence risk
and help estimating it, which will be the topic of this section. Through a critical review
of the presented privacy metrics we highlight and examine the assumptions that such risk
estimates are based on.

For clarity, we introduce an example:
We intend to publish differentially private statistics for the distance-work variable.
In \Cref{table:example} we provide an excerpt from that column
from the SOEP data set.
Note that the index column is only introduced to facilitate referencing.

\begin{table*}
	\caption{Exemplary data set of \textit{distance-work}.}
	\label{table:example}
	\begin{subtable}[t]{0.3\textwidth}
		\caption{Complete data set}
		\label{tab:complete}
		\centering
		\begin{tabularx}{0.6\textwidth}{Xr}
			\toprule
			id & distance \\
\midrule
			0 & 3 \\
			1 & 1\\
			2 & 10 \\
			3 & 675 \\
			4 & 17 \\
\bottomrule
		\end{tabularx}
	\end{subtable}
	\begin{subtable}[t]{0.35\textwidth}
		\caption{Neigbouring data set without maximum}
		\label{tab:neigbourmax}
		\centering
		\begin{tabularx}{0.5\textwidth}{Xr}
			\toprule
			id & distance \\
\midrule
			0 & 3 \\
			1 & 1\\
			2 & 10 \\
			3 & 675\\
			4 & 17 \\
\bottomrule
		\end{tabularx}
	\end{subtable}
	\hLine{-4.5}{-2.3}{-1.6}
	\begin{subtable}[t]{0.35\textwidth}
		\caption{Neigbouring data set without minimum}
		\label{tab:neigbourmin}
		\centering
		\begin{tabularx}{0.5\textwidth}{Xr}
			\toprule
			id & distance \\
\midrule
			0 & 3 \\
			1 & 1\\
			2 & 10 \\
			3 & 675 \\
			4 & 17 \\
\bottomrule
		\end{tabularx}
	\end{subtable}
	\hLine{13}{1.2}{15.9}
\end{table*}

\subsection{Assumptions of Risk Definition}
Even with the simple definition of global and local sensitivity,
subtleties arise regarding the
notion of \textit{neighbouring} databases that are decisive when calculating the sensitivity.
In some cases, neighbouring databases means that a
single value can be added or removed from a database to obtain a neighbouring
database. In other cases, it is assumed that a single value can be altered or
replaced, leaving the total number of entries unchanged. These different
interpretations are referred to as \textit{unbounded} and \textit{bounded}
differential privacy, respectively~\cite{lee_differential_2012}.

Lee and Clifton define
two databases as neighbours if one can be obtained from the other by adding or
removing exactly one individual of the data set.
The two databases therefore differ in size by exactly
one row and the smaller one is a subset of the larger one.

The model of Lee and Clifton assumes that sensitivities are computed
only between non-empty databases. 
However, it is entirely possible to construct filter queries that
return an empty data set. Therefore, it can be argued that global
sensitivity formulas should take empty data sets into account.
We argued that
global sensitivity, as the sole source of noise in our construction, is already
likely to be overly large. Only considering data sets with at least one member
can be seen as an attempt to limit global sensitivity.
While this decision is debatable, it is very unlikely that changing it would
have a strong effect on the overall risk patterns and the parameter
interactions we will point out in our~analysis.

\subsection{Sensitivity Calculation}
\label{sec:sensitivityknown}
The global sensitivity is not always known before data collection and must therefore often be estimated.
Even with a data set as large as the SOEP data set,
computing the \enquote{true} global sensitivity
of a query function remains conceptually impossible.
There can be no guarantee
that the most extreme values existing in the overall population
are present in the SOEP data set.
For some variables it is possible to determine
extreme values theoretically, for example if a
questionnaire item starts with `How many hours per day\ldots', guaranteeing
values between $0$ and $24$ after implausible survey responses are filtered
out.
Other variables, like the commuting distance in our example,
can vary almost indefinitely and have no specific bounds.
To still be able to calculate sensitivities,
we treat all values we have as our universe.
Distances between $1$ and $675$ become the practical
bounds for commuting distance, and based on these bounds
sensitivities are calculated individually per query type.

\subsubsection*{Mean}
Assuming that a valid data set always contains at least one entry, the largest
difference in mean between any two neighbouring databases is between one that
contains either the largest or the smallest value of the universe $U$, and one
that contains both. As a result, the global sensitivity of the mean query is
equal to half the distance between the largest and the smallest value of the
universe.

Referring to the example data in \Cref{table:example}, we are talking about the
values $1$ and $675$. Removing any of
the two from such a size $2$ database increases or decreases the \textit{mean}
value by $337$. This is the largest possible change between any two databases
from the universe \Cref{tab:complete}, representing the global sensitivity of
the \textit{mean} query.

For both the two-worlds and the many-worlds
risk, we also need to calculate local sensitivity for
a given data sample taken from the universe. It represents the largest
possible local influence any individual from the universe can have
on a query result calculated from this sample.

Local sensitivity for a given multiset $X$ from the universe $U$ 
is obtained by comparing four cases.
The \textit{mean} value of $X$ undergoes the largest local change by either
adding the largest or the smallest
value of the universe, or by removing the largest 
or the smallest value currently in the sample.
The largest absolute difference between these four cases from the initial \textit{mean}
value is equal to the local sensitivity $LS_{mean}(U, X)$.

Again referring to our example and considering
\Cref{tab:neigbourmax} as our sample $X$,
the \textit{mean} value changes from $7.75$
to $141.2$ by including the value $675$.
No larger difference in \textit{mean}
can be achieved by adding or removing any other value, resulting in a local
sensitivity of $133.45$ for data set \Cref{tab:neigbourmax} under universe \Cref{tab:complete}.

\subsubsection*{Median}
Similar to the mean, sensitivity of the median query is largest for databases
containing only the largest and the smallest value from the universe $U$.
Adding or removing a single value from a database, the largest possible jump in the
median function is between $\min(U)$ or $\max(U)$ and half the distance between $\min(U)$ and $\max(U)$, in any direction.

Local sensitivity of the median can be determined by shifting the current
median of a given multiset $X$ from the universe $U$ one position to the right
or to the left. This can be achieved by either adding/removing a value higher than the initial median, e.g. the largest value from
the universe, and then
determining the median of the resulting set.
Local sensitivity is equal to the
maximum absolute difference to the initial median between these two cases.
Adding or removing the smallest values from the universe or the subset,
respectively, would give an equivalent result.
With \Cref{tab:complete} our initial median is $10$.
With a sample like \Cref{tab:neigbourmax} we obtain a median of $6.5$ and with
\Cref{tab:neigbourmin} $13.5$, resulting in a maximal difference of $3.5$.

\subsubsection*{Min and Max}
Assuming that a valid database always contains at least one value, the largest
difference in the minimum/maximum value is equal to the distance between the smallest
and the largest value from the universe $U$. Given a database that only
contains the smallest value, adding the largest value leads to this largest
possible jump.

Given a multiset $X$ from the universe $U$, the minimum/maximum value can only be
influenced by either removing the smallest/largest value from the multiset, or by adding
a smaller/larger value from the universe $U$ that is not yet contained. In the latter
case, adding the smallest/largest value from the universe will maximize the difference.
Comparing the absolute differences of both cases gives the local sensitivity
for the max and min query.

\subsubsection*{Variance}

The variance of a distribution is defined as the mean squared difference from
the mean over all values of a sample.
The largest difference in variance between two neighbouring databases is
observed between one that contains two values, the largest and the smallest
from the universe $U$, and one where either of these two is removed. For the
first database the variance is at a global maximum for the given universe.
Adding any values in between the extremes or adding uneven amounts of the two
extreme values can only decrease the variance.
Thus, the global sensitivity is equal to half the squared
distance between the largest and smallest value of the universe.

\begin{table*}[t]
	\caption{Descriptive statistics of the numerical variables}
	\label{table:variables}
	\begin{tabularx}{\textwidth}{p{4cm}p{5.1cm}Xp{0.2cm}rrrr}
		\toprule
		variable (our label) & description & n & min & max & mean & median & std \\
		\midrule
		bjp\_05\_01 (work-hours) & Average hours spent on paid work / weekday
		& 25,238 & 0 & 18 & 5.35 & 7  & 4.38 \\
bjp\_82\_01 (distance-work) & Distance to workplace
		& 14,734 & 1 & 675 & 15.07         & 10    & 19.90 \\
bjp\_89\_01  (monthly-incomeGross)          & Gross monthly income
		& 16,594 & 0 & 2.7 M & 3,432.00 & 2,500    & 21,669.00 \\
\bottomrule
	\end{tabularx}
\end{table*}

Working with the example data from \Cref{table:example}, we obtain a sample
of maximum variance of $113569$ by including only the values $1$ and $675$.
Removing any of the two values gives a variance of $0$, leaving us with a global
sensitivity of the \textit{variance} query of $113569$ under universe \Cref{tab:complete}.

To calculate the local sensitivity, we proceed in the same way as for the other queries.
The variance should always be influenced the strongest by adding extreme values to
or removing them from a database. Given a multiset $X$ from the universe $U$,
we first compute the current variance of $X$. Then we either add one of the
extreme values from the universe or remove one of the extreme values from $X$.
Determining the largest absolute difference from the initial variance between
these four cases gives us the local sensitivity.
Regarding our example, the variance of \Cref{tab:neigbourmax} and of
\Cref{tab:neigbourmin} is $1838.75$ and $76737$, respectively.
Thus, the local sensitivity, as the maximum difference to our initial variance, is
$111730.25$.

\section{Empirical Evaluation of Risk Metrics}
\label{sec:evaluation}
The goal of our evaluation is to 
investigate the interaction of the privacy risk of Lee and
Clifton~\cite{lee_how_2011} with the SOEP data set.
Thus, we obtain
some initial insight into the concrete risk levels that could be communicated
to participants of scientific studies similar to the SOEP.

The entire analysis results can be
reproduced using the code in this repository~\footnote{\url{https://git.tu-berlin.de/j_allmann/master_thesis_code}},
This requires access to the SOEP data set, which is not included in the code
repository.

\subsection{SOEP Data Set and used Variables}
\label{sect:dataset}

For the SOEP study,
around 30,000 people in around 15,000 households have been and are
continued to be recruited across Germany.
Participants in the study are surveyed annually
to produce a large longitudinal data set.
Individuals in the data set are superficially anonymized, i.e.,
direct identifiers are removed and replaced with a pseudonym.
The data handling contract specifically forbids attempting to
re-identify individuals in the data set. This fact and other measures, like the
exclusion of precise geolocation data from the regular data set, show a basic
awareness of the dangers that re-identification attacks pose for such detailed,
superficially anonymized collections of sensitive individual data.

The SOEP data can be
considered highly susceptible to re-identification. It includes precise data
about the family status of a person in the past and the present, the size and
character of their home, their field of work and salary, including
changes over time, some self reported medical diagnoses, and other highly
individual and sensitive data.

The SOEP data set is not distributed as a single, unified database.
It consists of dozens of individual files, some containing raw survey data and
others containing preprocessed variables that are easier to handle than the raw data.
For our analysis, we work with a small subset of the variables of the main
SOEP questionnaire, limited to the year 2019,
being one of the most recent complete years in the SOEP at the time of writing.

The data set has $1159$ columns, the vast majority of which represent individual
items or sub-items from the survey questionnaire. Around $30$ columns contain
metadata such as participant and household IDs, biographical information, and
information about the interview procedure.
Study participants in the SOEP are assigned a random person ID (pseudonym) that allows
linkage of data across different tables, for example representing different
survey years. The total number of unique person IDs in our chosen data set is
$29905$ of $19085$ households
interviewed in the 2019 survey period.

With our analysis, we attempt to include variables spanning different
ranges, and statistical characteristics.
The chosen variables should be either likely to be
known to an attacker or considered sensitive to fit the narrative of an attacker
aiming to identify an individual and subsequently learn sensitive facts about them.
As a result, we selected the variables listed in \Cref{table:variables}.
Our selected variables are numerical,
since the majority of query types we examine only make sense on numerical
variables.

\subsection{Procedure}\label{sect:procedure}
For each variable, listed in~\Cref{table:variables}, we estimate the
two-worlds risk $\rho_{tw}$ and compare it to the
many-worlds risk $\rho_{mw}$ and global risk $\rho_{gl}$.
First, negative values representing missing values in the SOEP
data set are removed.
Obviously, each variable
already represents a sample taken from the much larger universe that is the
general population. 
However, for our purpose we treat each variable as its own
universe, since for us these are all the values we know can exist.
We treat all available positive values as
our data universe as explained in~\Cref{sec:sensitivityknown}.
Thus we use the maximum value in the data set as the global maximum.
The global minimum value is set to $0$
if it is a valid answer for the specific variable.

We calculate the risk for all our presented query types with different
values of $\varepsilon$.
To evaluate the influence of the data characteristics, we 
iterate over a set of sample sizes that are taken from
the universe of each variable.
Since each variable contains a different
number of valid values, and they differ substantially in size, the size of the
subsets is different for each variable.
Due to its randomness, we repeat each experiment 100 times.

Global sensitivity is based on the whole
variable representing the universe, local sensitivity is computed from the
respective sample and the universe as explained in \Cref{sec:sensitivityknown}.
We compute the many-worlds risk $\rho_{mw}$,
the two-worlds risk~$\rho_{tw}$ and the global risk~$\rho_{gl}$ that does not factor in the sample size.

\subsection{Results}\label{sect:results}

\begin{figure*}[tb]
	\begin{minipage}[t]{\columnwidth}
			\includegraphics[width=\columnwidth]{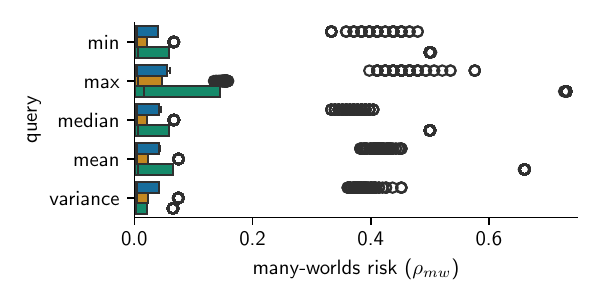}
\end{minipage}
	\begin{minipage}[t]{\columnwidth}
		\includegraphics[width=\columnwidth]{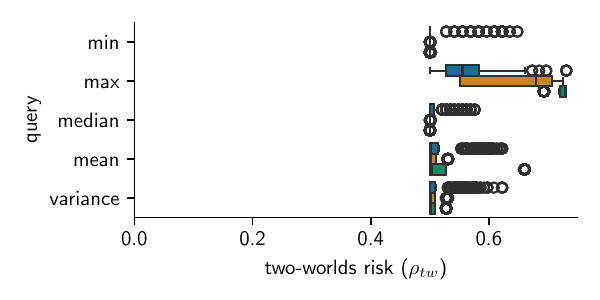}
\end{minipage}
	\begin{minipage}[t]{2\columnwidth}
		\centering
		\vspace*{-0.5cm}
		\includegraphics[width=0.5\columnwidth]{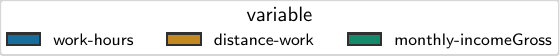}
	\end{minipage}
	\begin{minipage}[t]{\textwidth}
		\captionsetup[sub]{labelformat=parens}
		\begingroup
		\captionsetup{type=figure}
		\begin{subfigure}{0.5\columnwidth}
			\vspace*{0.5cm}
			\caption{Many-worlds risk.}
			\label{figure:plot_many_worlds_box}
		\end{subfigure}
		\begin{subfigure}{0.5\columnwidth}
			\vspace*{0.5cm}
			\caption{Two-worlds risk.}
			\label{figure:plot_two_worlds_box}
		\end{subfigure}
		\endgroup
	\end{minipage}
	\begin{minipage}[t]{\textwidth}
		\caption{Comparison of many-worlds risk and two-worlds risk with $\varepsilon=1$.}
		\label{fig:risk_metrics}
	\end{minipage}
	\vspace*{-1em}
\end{figure*}

\subsubsection*{Comparison of Risk Metrics}

The differences in risk between the many-worlds metric ($\rho_{mw}$)
and the two-worlds risk ($\rho_{tw}$) are striking. 
While the two-worlds risk can never go below
$0.5$, the risk computed with the many-worlds formula can become arbitrarily
small with increasing sample size.
Comparing these two metrics is not a question of which one is better, but which
one is easier justified under a specific interpretation of the definition of
differential privacy.

Small risks resulting from the assumptions underlying the many-worlds 
metric give a false impression of safety as long as the
sample size is large enough. The two-worlds metric captures the
assumption that only the presence or absence of a single individual is unknown
to an attacker. This fact can always be guessed, and the risk of identification
for an individual is always between $0.5$ and $1$, the latter meaning that an
attacker can determine the presence or absence of an individual with absolute
certainty.

We can directly compare $\rho_{mw}$ and $\rho_{tw}$.
Comparing these the two metrics in
\Cref{fig:risk_metrics}
it becomes clear, how important it is to thoroughly review
assumptions and definitions underlying a risk metric to arrive at meaningful
and comparable results. While the differences in the construction of the two
metrics are subtle,
the resulting numbers are worlds apart. Any attempt at successfully educating
users about differential privacy will have to deal with these ambiguities.
Therefore, we consider the $\rho_{tw}$ to be more accurate.

The worst-case risk ($\rho_{gl}$), as already mentioned
in~\cite{mehner_towards_2021}, represents a robust upper bound for the tracing
risk of a differentially private system.
With $\varepsilon=1$, the worst-case risk results in $\rho_{gl}=0.731$.
This metric, with its advantage
of relying only on $\varepsilon$ and not requiring any knowledge of the data
universe, can be considered a very useful tool under certain circumstances.

\subsubsection*{Influence of Query Type}
The effects of the query type are also visible in \Cref{figure:plot_two_worlds_box}.
Regardless of the variable, the risks for the query \textit{max} varies more.
The \textit{max} query shows a
complex pattern of risk values
overall spanning the whole range of possible risks.
The \textit{min} query,
however, stays in average below a risk of $\rho_{tw} = 0.51$ for the variable \textit{work-hours}.
For the other two variables the risk is around $0.5$.
This contrast is especially
surprising as \textit{min} and \textit{max} appear to be very similar queries,
operating symmetrically on opposite ends of a sample distribution. The reason
for this observation then must lie in an interaction with the underlying data.
Consulting~\Cref{table:variables}, the variable \textit{distance-work}
captures participant's commuting distance.
As can be expected intuitively, it
is a rather bottom heavy distribution with outliers at the upper end. Most
people tend to live close to their workplace, and only few travel very far.
Under these conditions, it is likely that even in small samples at least some
very small values are present and local sensitivity of the \textit{min} query
is small. At the same time, local sensitivity for the \textit{max} query can
frequently be large, when none of the large outlier values is present in a
sample. Any of the outliers entering such a database is at high risk of being
identified.

\subsubsection*{Influence of Sample Size}
\begin{figure*}[tb]
	\begin{minipage}[t]{\columnwidth}
		\centering
		\includegraphics[width=0.9\columnwidth]{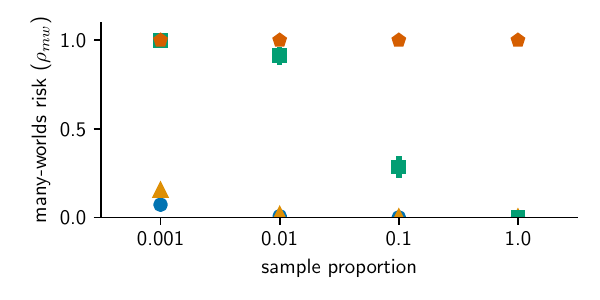}
	\end{minipage}
	\begin{minipage}[t]{\columnwidth}
		\centering
		\includegraphics[width=0.9\columnwidth]{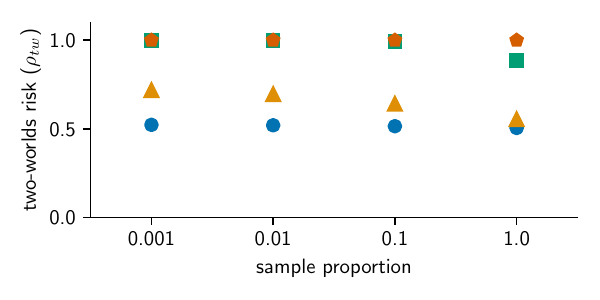}
	\end{minipage}
	\begin{minipage}[t]{2\columnwidth}
		\centering
		\vspace*{-0.7cm}
		\includegraphics[width=0.5\columnwidth]{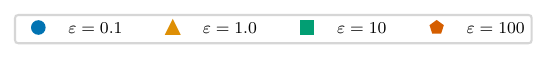}
	\end{minipage}
	\begin{minipage}[t]{\textwidth}
		\captionsetup[sub]{labelformat=parens}
		\begingroup
		\captionsetup{type=figure}
		\begin{subfigure}{0.5\columnwidth}
\caption{Many-worlds risk}
			\label{figure:plot_many_worlds_sample}
		\end{subfigure}
		\begin{subfigure}{0.5\columnwidth}
\caption{Two-worlds risk}
			\label{figure:plot_two_worlds_sample}
		\end{subfigure}
		\endgroup
	\end{minipage}
	\begin{minipage}[t]{\textwidth}
		\caption{Risks with varying sample proportion and $\varepsilon$ for \textit{max} query and variable \textit{distance work}.}
		\label{fig:sample_prop}
	\end{minipage}
	\vspace*{-1em}
\end{figure*}

In \Cref{fig:sample_prop} we provide the risks for our selected variables with
the sample proportion on the x-axis for the \textit{max} query.

The influence of sample size on risk unfolds in two different ways, depending
on the risk metric that is observed. For the many-worlds metric, sample size
has a direct and drastic effect. Under the assumptions of this metric, the
number of possible worlds an attacker has to consider, depends on the number of
entries in a database. More possible worlds result in a smaller overall risk
for each individual, an effect represented by the factor $n$ in the denominator
of the risk \Cref{formula:full_risk_metric}.
This is visible in \Cref{figure:plot_many_worlds_sample} for $\varepsilon=10$.
Very small sample sizes, $\rho_{mw}$ approaches the worst case of $1$,
while large sample sizes keep this risk at
values below $\rho_{mw} = 0.01$ even for a large value $\varepsilon = 10$.

In \Cref{figure:plot_two_worlds_sample} we see that $\rho_{tw}$ only decreases slightly, as
the influence of sample size is only indirect. As in
this case only two possible worlds are considered, the latter does not directly factor into $\rho_{tw}$.
An indirect effect can be observed
through a connection between sample size and local sensitivity. Most queries
are less locally sensitive on a larger sample, because query functions like
\textit{mean}, \textit{variance}, or \textit{median} are intuitively more stable
in larger samples if only few individuals change.
A similar observation can be made for the \textit{min} and \textit{max} queries.
In larger samples, it becomes less likely that the minimum or maximum value in the sample
and, respectively, the minimum or maximum value in the universe are far apart,
which would result in high local sensitivity.

In summary, sample size has a strong effect on risk under many conditions. It
is most striking for the many-worlds risk metric, where it is an explicit
component of the formula and leads to risks that are orders of magnitude smaller
than even theoretically possible under the two-worlds metric.
For the latter,
sample size affects risk indirectly through its influence on local sensitivity.

\subsubsection*{Influence of Epsilon}
The privacy parameter $\varepsilon$ has the most direct influence on
risk in differential privacy. It is the central parameter balancing privacy
against utility.
Generally, a larger $\varepsilon$ allows for larger differences between two
neighbouring databases, which reduces the required amount of noise.
As a direct
result, the risk of being identified increases for each data subject, by
allowing an attacker a larger knowledge gain through each noisy query result
they obtain.
This effect is visible in \Cref{fig:sample_prop}.
Under certain conditions, epsilon becomes the only parameter influencing risk.
In cases where local and global sensitivity are equal, the risk metric is
reduced to the global risk metric taking $\varepsilon$ as its only parameter.

There is also an interaction with $\varepsilon$ and the sample proportion. 
Sample size only plays a role in a~certain region
of $\varepsilon$ values, while losing its effect at very small and at very large
$\varepsilon$, where risk generally approaches the minimum or maximum values of
$\rho_{tw} = 0.5$ or $\rho_{tw} =1$,~respectively.

\subsubsection*{Summary of Findings}
The privacy parameter $\varepsilon$ has the strongest and
most predictable effect on risk.
The effects of
other parameters are usually most strongly pronounced in this critical window,
while they tend to vanish for very large or very small $\varepsilon$.

The magnitude of the sample size
effect is harder to specify than it is for $\varepsilon$. In general, this means that reliable risks could only
be communicated to the data subjects of a differentially private system if the
number of participants can be estimated in advance. Even then, queries
filtering out small subsets of users would present greater risks than
statistics computed over the entire user base of a system.

A robust risk value can only be computed if sample sizes are
guaranteed to be at least of a certain size. While we refrain from quantifying
such a lower bound, this requirement for computing reliable risk metrics
appears manageable.

\section{Discussion}
\label{sec:discussion}
The overarching goal of our case study is to show how a differential privacy
risk metric is under real world conditions.
We discuss our results from two angles.
Firstly, we reflect on the stability of the risk metric, and
secondly, we discuss the general suitability of using the proposed metric to communicate the privacy guarantees.

\subsection{Stability of the Risk Metric}
\subsubsection*{Under changing calibrating noise}
Local sensitivity
often allows adding much smaller amounts of noise, especially for larger
numbers of participants~\cite{nissim_smooth_2007}.
This directly leads to an
increase in query precision and therefore in data utility. 
However, as Nissim et al. pointed out~\cite{nissim_smooth_2007},
local sensitivity itself leaks information
about the values included in the data set. The authors propose a remedy for this
problem in the form of smooth sensitivity. Smooth sensitivity attempts to
compute a smooth upper bound for local sensitivity, ensuring that the
information leaked by the local sensitivity computation itself is limited.

At the centre of the Lee and Clifton risk formula, however, and in turn also of
our modified risk formula, lies the assumption that noise is simply calibrated
to global sensitivity. The fraction between global and local sensitivity
crucially influences the risks computed with both metrics. If noise were to be
calibrated to (smooth) local sensitivity, this fraction would approach or
become $1$, leaving us with the worst-case global privacy
risk~\ref{formula:global_privacy_risk}. This is very important to note, since
it directly connects risk and utility optimization in an opposing manner. It
also reinforces the notion that the worst-case metric, however crude it may
appear, might be the most viable metric in practical application scenarios.

\subsubsection*{For composite queries}
Data analysis rarely consists in one single, isolated query. The risks depicted
in our plots represent risks for single, simple queries against single,
isolated variables. In reality, data analysts usually query a data set many
times, using composite queries that usually involve more than one variable. One
of the very elegant features of differential privacy is the composition
theorem, which states, that generally the $\varepsilon$-values of repeated queries
add up~\cite{kairouz_composition_2017}.
It can
be assumed that the risks calculated here behave similarly, and that a risk
budget granted to a data analyst is a viable concept.
If a certain risk budget was allocated and
communicated to end users, it would be important for an analyst to know how
strongly different queries and combinations of queries use up this budget. Our
results can help inform practical methods to compute such information in
realistic scenarios.

\subsection{Suitability for Communicating Risks}
The effects of the different parameters on risk could be demonstrated
and analysed in previously not achieved levels of detail.
Detailed risk metrics could prove useful in
situations where the necessary parameters are known, while worst case estimates
could serve as a fallback whenever this is not the case. How to turn these
different metrics into viable information formats to communicate the risk is an open
question.

The notion of a 50\%
privacy risk before even making any decision to participate or not, as implied
by the two-worlds metric, is already not very intuitive~\cite{franzen_am_2022}. Successfully
conveying the fact that subtle differences in the construction of the metric
can lead to risks in the order of fractions of a percent, seems unrealistic at
best. A road out of this dilemma might lie in using the metrics only for
comparing different systems, instead of trying to communicate absolute risks.
Being able to state that one system has a less risky parameter set than a
competing one appears possible even despite the complex nature of the risk
metrics that became apparent in our analysis.

\section{Related Work}
\label{sec:related_work}

While there has been research on communicating the mechanisms behind
differential privacy
\cite{franzen_am_2022,wood_differential_2018,cummings_i_2021,nanayakkara_visualizing_2022,xiong_using_2022}, the results are still scarce.
Initial findings so far have
indicated that the success of these explanations partly depends on prior
knowledge and numeracy of participants, indicating the need for adaptive
explanation formats~\cite{franzen_am_2022}.

Cummings et al.~\cite{cummings_i_2021} have investigated the interaction
between users' specific expectations towards a differentially private system
and the way differential privacy is described, regarding their willingness to
share data with the respective system. Results indicate that a precise
description of differential privacy might not be enough to increase users'
motivation to share data. Instead, the extent to which their individual privacy
concerns can be addressed by differential privacy and the extent to which this
is stressed in the description have a major influence on sharing behaviour.

Xiong
et al.~\cite{xiong_using_2022} have made initial contributions to studying
visual explanation formats. Their approach focuses on visualization of data
utility and puts less emphasis on the visualization of remaining privacy risks,
the opposite side of the privacy utility trade-off. Additionally, interactive
communication designs are pointed out as a promising approach to investigate in
the future.

For data subjects,
risk can be assumed to be of larger
importance than data utility.
Unfortunately, quantifying these
risks under realistic conditions has not been a research focus so far. Early
work on concrete risks was published by Lee and Clifton~\cite{lee_how_2011},
and it was later picked up in the context of risk
communication~\cite{franzen_am_2022}. The coverage of this topic still appears
unsatisfying.

While there are many case studies in which the re-identification risk of a large data set is determined~\cite{dankar2012estimating},
in the area of differential privacy, this is rather limited.
Example scenarios often consist of very small, artificial databases with few
individuals and few variables that suffice to illustrate the mechanisms of
differential privacy. These toy scenarios, however, often do not allow
reasoning about the performance of a differentially private system under real
world conditions.

Theoretical research often avoids discussing concrete parameter choices and
points out, that $\varepsilon$, which balances
privacy and utility against each other, needs to be chosen depending on the
application context and the actual data
involved~\cite{dwork_differential_2019}. Commercial actors, on the other hand,
tend to be rather secretive about the exact implementation decisions and
parameter choices of their systems. When such information is released, be it in
research or in commercial contexts, a huge variance of presumably reasonable
parameter choices emerges. This issue has been pointed out by Dwork et
al.~\cite{dwork_differential_2019}, leading to the proposal of a public
registry of such parameter choices to guide decision-making for future
differential privacy applications.

In this context, reasoning about concrete individual risks forms an important
part of the overall picture.
This is where we pick up the thread in this paper and intend
to narrow existing knowledge gaps about risks under differential privacy in
realistic scenarios.

\section{Conclusion and Outlook}
\label{sec:conclusion}

In this paper, we conduct a case study to estimate
risk metrics for differential privacy
of an SOEP data set.
Computing robust, meaningful
risk estimates for a differentially private system as a whole is challenging.
We discuss an existing differentially identifiability risk metric and
highlight the need of some requirements, e.g.,
that an attacker can always guess the
presence or absence of an individual with a probability of 50\%.
This metric, similar to its predecessors, depends on several factors,
including the privacy parameter $\varepsilon$, the data that is handled, and the
query that is applied to the data.
Especially our evaluation with the SOEP data set shows that
the influence of data characteristics and query type
make it hard to obtain reliable risk values to communicate the risk beforehand.
Falling back to the worst-case risk metric depending
only on $\varepsilon$ might be the only practical option here.

Our results can help form a more detailed picture of the performance of
differentially private systems, both for professionals as for untrained end
users.
Identifying particularly queries with higher risks and particularly vulnerable attributes
should also become easier building on the results and the methodology presented
here. 
This could lead to more fine-grained calibration techniques for complex
analyses, possibly increasing data utility while keeping data subjects better
informed about how strongly their privacy is affected.
Computing risks and situating them in real-world
scenarios is hard, but with the results presented here it should have become
easier.

\bibliographystyle{plain}

\appendices

\end{document}